# Towards Structure-Aware Surrogate Modeling: Explicit Region Interaction Improves Knee Contact Stress Prediction

Zhengye Pan[a,b], Jianwei Zuo[a,b], Jiajia Luo[*a,b]

[a] Biomedical Engineering Department, Institute of Advanced Clinical Medicine, Peking University, Beijing, China

[b] Institute of Medical Technology, Peking University Health Science Center, Peking University, Beijing, China

[*] Corresponding author

**Abstract**

Knee contact-stress hotspots are closely linked to meniscal/cartilage injury risk. Still, high-fidelity subject-specific FEA is too computationally expensive for large-cohort, multi-condition, near-real-time use. Existing MeshGraphNet-style surrogates mainly rely on stacked local message passing, which is often insufficient for modeling long-range dependencies and limits interpretability. This study benchmarked a deep-stacked baseline model against three explicit region-interaction architectures. Using a 90° change-of-direction task and a strict cross-subject evaluation framework, we assessed whole-field error, peak stress fidelity, and hotspot spatial consistency under matched computational budgets. Region-interaction models significantly reduced whole-field nodal stress errors compared to the purely stacked baseline. Crucially, they achieved markedly higher accuracy in reconstructing the high-stress tail and demonstrated superior spatial consistency and temporal robustness in localizing high-risk stress hotspots. Explicit region-level interaction provides a more structure-aligned surrogate modeling paradigm for knee contact mechanics and yields stronger risk-relevant stress phenotype recovery under comparable computational budgets, while supporting more interpretable injury-risk assessment.

# 1 Introduction

The knee experiences complex, multi-tissue coupled contact and load transfer during dynamic tasks such as gait, jumping, and change-of-direction maneuvers. The resulting distribution of intra-articular contact stress—especially the formation and migration of localized high-stress hotspots—is considered closely related to the risk of meniscal and cartilage injury [1,2]. Subject-specific finite element analysis (FEA), built on individual anatomy and boundary conditions, can characterize contact stress and strain fields at the tissue and mesh scales and is an important tool for quantitatively studying joint contact mechanics [3]. However, high-fidelity FEA typically requires extensive preprocessing and iterative numerical solution, resulting in high computational cost and making it difficult to support large cohorts, multiple conditions, and near-real-time analysis needs [4,5]. This, in turn, limits broader adoption in rapid assessment, risk warning, and clinical applications.

In recent years, deep surrogate models have offered a feasible path for near-real-time prediction of joint contact mechanics. Graph neural networks (GNNs), which represent finite element meshes as graphs, have attracted attention because they naturally accommodate irregular topologies [4]. MeshGraphNet (MGN)-style GNNs perform layer-by-layer aggregation of neighborhood information through node-level local message passing, enabling regression of full-field mechanical quantities [6–8]. The implicit core assumption is that stacking more local propagation layers (or increasing width) can progressively expand the receptive field and thereby approximate the joint's true nonlocal load transfer and stress redistribution. But in a multi-tissue contact system such as the knee, true propagation may not require uniform participation of the entire mesh: key information that drives hotspot formation may be dominated by coupling among only a small number of load-bearing/contact-relevant regions [9], while many other nodes mainly reflect passive responses. Meanwhile, relying solely on deeper/wider networks to strengthen cross-region dependency modeling incurs rapidly increasing memory and compute cost [10,11]. It makes it difficult to answer mechanism questions in a structured way—such as which regions dominate propagation and how regions interact—creating bottlenecks in both efficiency and interpretability.

Against this background, this study focuses on a more fundamental and testable scientific question: Is intra-articular stress propagation governed mainly by interactions among a small number of functional (load-bearing) regions? If this hypothesis holds, we can explicitly introduce region-level interactions into the model architecture, expressing long-range dependencies in a lower-dimensional and more structured way. This could improve hotspot-relevant prediction quality without substantially increasing computational budget and could yield interpretable region-coupling phenotypes. We therefore propose and test the following structural hypothesis: compared with approaches that rely only on stacking local propagation, architectures that explicitly model region-level interactions are better suited to capturing long-range dependencies in knee contact stress. We expect more substantial gains in hotspot consistency and in controlling high-quantile errors, while still achieving stable improvements in whole-field error.

To verify this hypothesis and enable a clear paradigm comparison, we adopt a strict cross-subject evaluation in which test participants are completely held out. We use a rapid change-of-direction (CoD) maneuver—where high-stress risk is particularly prominent—as the representative task [12,13]. Using a deep-stacked local-propagation baseline model (MGN-Stack) as the reference, and under matched training budgets and GPU memory constraints, we benchmark it against three region-interaction designs with clearly distinct structure: Region-Interaction MGN (RI-MGN), Hierarchical MGN (H-MGN), and Region-gated MGN (RG-MGN). Beyond whole-field error, we further emphasize hotspot consistency and high-quantile stress error—metrics that are more sensitive to long-range dependencies—as primary comparisons, thereby more directly testing whether explicit region interaction can provide a stronger inductive bias for modeling knee contact mechanics. This, in turn, supports a shift in modeling paradigm from expanding the receptive field purely by stacking toward structure-aware region-interaction modeling.

## 2 Methods

**2.1 Dataset Construction**

2.1.1 Participant Recruitment and Data Collection

This study recruited nine adult male soccer players (174.4 ± 4.3 cm, 72.43 ± 7.1 kg, 22.1 ± 1.7 years). Inclusion criteria required more than seven total years of soccer participation.

Participants were also required to have no history of knee pain or prior knee surgery, no evident lower-limb functional impairment, and a dominant right leg. Participants were instructed not to compete within 48 hours before testing. Each participant was informed of the study purpose and procedures before the experiment, and all participants signed written informed consent.

Kinematic data were collected using a Vicon 3D motion-capture system (200 Hz, V5, Oxford, United Kingdom). Ground reaction force (GRF) data were synchronously recorded using a Kistler 3D force plate (1000 Hz, 9286AA, Winterthur, Switzerland). The reflective marker placement followed the setup described by Pan et al. [14]. Before formal testing, each participant completed a 5-minute warm-up including jogging (2.5 m/s) and self-selected stretching. During formal testing, participants were asked to perform five 90° CoD trials (Fig. 1) from the same starting posture at an approach speed of $5 \pm 0.5$ m/s. Of these, three trials were selected for subsequent analysis: those with approach speeds closest to the target value and with full-foot contact on the force plate. An infrared speed-timing gate was placed 5 m from the center of the force plate to monitor approach speed; the gate height was set to match the participant's greater trochanter height. Initial contact was defined as vGRF > 20 N, and toe-off was defined as vGRF < 20 N; the interval between the two was defined as the stance phase.

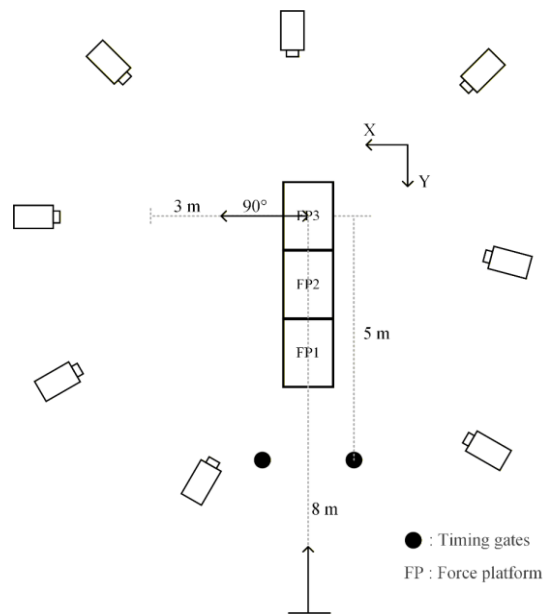

Figure 1. Experimental setup of the 90° change-of-direction task. Participants sprinted along an 8 m approach, stepped on three sequential force platforms (FP1–FP3), and performed a 90° cut into a 3 m lateral lane. Timing gates (black dots) were used to monitor approach speed, and X/Y indicate the laboratory coordinate axes.

### 2.1.2 Data Processing

A fourth-order, bidirectional, zero-phase Butterworth low-pass filter with a 10 Hz cutoff frequency was applied to filter the measured marker trajectories and GRF signals. Using the Rajagopal musculoskeletal model configured in the OpenSim (v4.5, Stanford University, USA) environment, the knee joint was modeled with three degrees of freedom—flexion/extension, internal/external rotation, and abduction/adduction—and inverse kinematics and joint reaction analysis were performed [15]. Knee spatial pose and loading were described in a unified tibial anatomical coordinate system, which provided standardized driving boundary conditions for subsequent mechanical simulations.

### 2.1.3 Finite Element Simulation and Dataset Generation

Finite element models were constructed based on the standardized topology mesh from OpenKnee(s) [16]. Quasi-static simulations were performed in FEBio with a time step of 0.01. Cartilage, ligaments, and menisci were assigned isotropic and transversely isotropic Mooney–Rivlin hyperelastic properties, and ligament pre-tension was included. For boundary conditions, knee kinematics computed in OpenSim were applied as kinematic constraints, and joint reaction forces were applied as equivalent loads to a distal femur reference node.

Each finite-element time step was packaged as a graph-structured sample to generate the dataset for training the surrogate model. Nodes were defined at element centroids; node features included initial geometric coordinates and a global vector that fused time-varying pose and load information. For von Mises stress outputs, a combined strategy of logarithmic transformation and z-score normalization was applied to balance the numerical distribution and improve GNN training performance.

## 2.2 Deep Surrogate Models

### 2.2.1 Network Architecture

To test the hypothesis above, this study designed four network architectures that share the same MGN backbone and adopt an encoder–processor–decoder paradigm. Node/edge encoders map inputs to latent representations; the processor performs local message passing across the mesh and iteratively updates node representations; the decoder outputs element-level stress. To ensure comparability, the four models share identical input/output definitions and encoder/decoder designs; differences lie only in how the processors represent long-range dependencies.

First, MGN-Stack is used as the baseline model. It introduces no additional global module and expands the receptive field solely by increasing the number of local message-passing steps $T$. This model serves as a baseline for whether simply stacking local propagation can approximate the nonlocal load transfer and stress redistribution processes inside the knee. RI-MGN introduces a small number of region tokens (virtual nodes) as a low-dimensional channel for region-level interactions. Specifically, full-graph node representations are softly assigned and weighted, aggregated into K tokens; interactions are performed in token space; and the post-interaction token information is injected back into nodes and fused with local message-passing results. The core idea is that if long-range stress dependencies are driven mainly by interactions among a few functional regions, token-based region coupling can improve long-range representation with low incremental cost.

H-MGN explicitly builds coarse-scale representations via a graph-hierarchical mechanism: nodes are aggregated into fewer supernodes, interactions are performed on the coarse graph, and coarse information is mapped back to the original nodes and fused. In contrast to RI-MGN's token bottleneck, H-MGN emphasizes multi-scale structure—modeling long-range dependencies at a lower resolution and then feeding them back to refine the fine-scale field. RG-MGN does not add an extra broadcast-back branch; instead, it encodes region coupling directly in the message-passing operator. The model predicts implicit region assignments and uses them to generate edge-wise gating coefficients that modulate message gains, allowing propagation strength to vary—learnably and in a structured way—across different region pairs. This corresponds to a stronger attribution hypothesis: long-range effects are reflected in the operator itself (via region-dependent propagation) rather than compensated through an additional global-aggregation path.

To ensure that conclusions can be attributed to long-range structural representations, the four models were kept consistent in input features, supervision signals, backbone form, and training configurations. Only key structural hyperparameters were adjusted so that all models operated under the same GPU memory budget, enabling a comparison between stacking local propagation and explicit region interaction in terms of both accuracy and efficiency.

### 2.2.2 Training and Inference

Grouped 3-fold cross-validation was used. The nine participants were grouped by ID into three folds (Fold 1: $P_1$-$P_3$; Fold 2: $P_4$-$P_6$; Fold 3: $P_7$-$P_9$). In each fold, models were trained on data from two folds and tested on the remaining fold, with participants completely unseen. For each cross-validation fold, all normalization parameters (including the log-transform setting and z-score statistics) were estimated using the training participants only and then applied unchanged to the test participants, thereby preventing data leakage. To avoid regression instability caused by invalid elements, only nodes marked as valid by a mask were included when computing errors during training and validation, using masked MSE as the objective function:

$$\mathcal{L}(\theta) = \frac{1}{\sum_i m_i} \sum_i m_i \parallel \hat{y}_i - y_i \parallel_2^2 \qquad (1)$$

where $m_i \in \{0,1\}$ is the node mask, and $y_i$ and $\hat{y}_i$ are the normalized stress label and prediction, respectively. During evaluation and visualization, model outputs were transformed back to physical units using the inverse of the transforms used during training. Except for structural hyperparameters, all other training settings were kept consistent across models. Parameter updates used the AdamW optimizer, and a Reduce-on-plateau learning-rate scheduler (initial learning rate $1\times10^{-3}$) was used based on validation loss; early stopping was also used to mitigate overfitting. Given the graph sample size and GPU memory constraints, training used small batches and applied gradient-norm clipping to stabilize convergence. All four models enabled automatic mixed precision (AMP; mixed-precision arithmetic) to improve training throughput [17]. To keep model capacity comparable, all four models had the same backbone hidden dimension and message-passing depth (Hiddens = 32), and processor parameters were shared across iterative steps. To characterize the long-range capability of pure stacking, MGN-Stack used two propagation-step settings (MGN3: steps = 3; MGN6: steps = 6), whereas the other models kept steps = 3 and captured long-range dependencies via their region-interaction mechanisms. All models were trained on four NVIDIA TITAN GPUs (batch size = 1; epoch = 120).

### 2.2.3 Evaluation Metrics

To comprehensively evaluate each model's reconstruction of stress fields and highlight sensitivity to long-range dependencies/high-risk-region errors, we compared models across three perspectives: whole-field error, peak/high-quantile fidelity, and spatial consistency.

For whole-field error, root mean square error (RMSE) and mean absolute error (MAE) quantified overall prediction deviation, and normalized RMSE (nRMSE) was reported to reduce unit dependence. Let $N$ be the number of nodes at time step $t$, with FEM ground truth $y_i$ and model prediction $\hat{y}_i$ at node $i$.

$$\text{RMSE} = \sqrt{\frac{1}{N}\sum_{i=1}^{N}(\hat{y}_i - y_i)^2} \tag{2}$$

$$\text{MAE} = \frac{1}{N}\sum_{i=1}^{N}|\hat{y}_i - y_i| \tag{3}$$

$$\text{nRMSE} = \frac{\text{RMSE}}{\max_j(y_j)} \tag{4}$$

where nRMSE is normalized by the maximum ground-truth stress max(y) at the current time step, reflecting prediction error as a proportion of the instantaneous peak load.

To evaluate high-stress concentration regions, we define peak relative error $RE_{\max}$ and 95th-percentile relative error $RE_{P95}$ to specifically test the "peak-shaving" phenomenon commonly seen in deep surrogate models:

$$RE_{max} = \frac{|max_i(\hat{y}_i) - max_i(y_i)|}{max_i(y_i)} \tag{5}$$

$$RE_{P95} = \frac{|P_{95}(\hat{y}) - P_{95}(y)|}{P_{95}(y)} \tag{6}$$

where $P95(\cdot)$ denotes the 95th percentile. These two metrics capture prediction bias in the highest-stress tail (often corresponding to the most hazardous meniscal region). To evaluate whether models spatially localize high-stress risk regions correctly, we compute node-wise Pearson correlation ($r$), and introduce Dice coefficient and intersection-over-union (IoU) metrics to evaluate spatial overlap of high-risk regions:

$$r = \frac{\sum(\hat{y}_i - \bar{\hat{y}})(y_i - \bar{y})}{\sqrt{\sum(\hat{y}_i - \bar{\hat{y}})^2}\sqrt{\sum(y_i - \bar{y})^2}} \tag{7}$$

$$\text{Dice} = \frac{2|S_{true} \cap S_{pred}|}{|S_{true}| + |S_{pred}|} \tag{8}$$

$$\text{IoU} = \frac{|S_{pred} \cap S_{true}|}{|S_{pred} \cup S_{true}|} \tag{9}$$

where H is the set of high-risk-region nodes whose stress exceeds the 95th percentile, and Ĥ is the corresponding predicted set. Further, to quantify hotspot localization ability, we introduce a hotspot-distance metric $d_t^{hot}$ to measure the spatial offset between ground-truth and predicted high-risk regions. Let $\mathbf{p}_i$ denote the centroid coordinates of node $i$. The geometric centers of the ground-truth and predicted high-risk regions are:

$$\mathbf{c}_t^{true} = \frac{1}{|S_t^{true}|}\sum_{i \in S_t^{true}} \mathbf{p}_i \tag{10}$$

$$\mathbf{c}_t^{pred} = \frac{1}{|S_t^{pred}|}\sum_{i \in S_t^{pred}} \mathbf{p}_i \tag{11}$$

and the hotspot distance is defined as:

$$d_t^{hot} = \| \mathbf{c}_t^{true} - \mathbf{c}_t^{pred} \|_2 \tag{12}$$

### 2.3 Statistical Analysis

A nonparametric statistical testing framework was used to compare model performance across metrics while treating the participant as the primary inferential unit. First, the Friedman test was used to examine overall differences among models in global error, peak fidelity, and spatial consistency. When the overall effect was significant, paired Wilcoxon signed-rank tests were used for post hoc pairwise comparisons, and the Holm–Bonferroni procedure was applied to correct for multiple comparisons. A two-sided significance level $\alpha = 0.05$ was used. Metrics are reported as mean ± standard deviation (Mean ± SD), and statistical analysis was performed in Python (v3.9.13, Delaware, US).

## 3 Results

### 3.1 Whole-Field Nodal Stress Reconstruction Accuracy

Overall error performance of the five models for whole-field nodal stress reconstruction is shown in Table 1. Significant differences were observed in RMSE and MAE across models ($p < 0.05$). MGN3, which relies on pure stacked local propagation, had the largest error. Although increasing the propagation steps to 6 (MGN6) significantly reduced error, it remained noticeably worse than the three models with explicit regional interactions. Under the same training budget, RI-MGN, H-MGN, and RG-MGN further and significantly reduced whole-field error. A stable performance stratification was observed among the region-interaction models: H-MGN achieved the lowest overall error and significantly outperformed RI-MGN and RG-MGN; RI-MGN also significantly outperformed RG-MGN.

Table 1. Comparison of whole-field nodal stress prediction errors across models (mean ± SD)

|      | MGN3          | MGN6         | RI-MGN         | H-MGN            | RG-MGN            |
|------|---------------|--------------|----------------|------------------|-------------------|
| RMSE | 0.487 ± 0.055 | 0.143 ± 0.02 | 0.062 ± 0.006  | **0.058 ± 0.007** | 0.065 ± 0.007    |
| MAE  | 0.375 ± 0.03  | 0.108 ± 0.02 | 0.04 ± 0.004   | 0.042 ± 0.005    | **0.037 ± 0.005** |

Note: **Boldface** indicates results that are significantly different from all other models within the same metric.

Whole-field reconstruction consistency and normalized-error characteristics are shown in Figure 2. MGN3 had the lowest Pearson correlation coefficient and the largest dispersion, and the highest nRMSE. Increasing propagation steps to 6 significantly improved consistency and reduced nRMSE. However, compared with the three explicit region-interaction models, MGN6 still showed a significant gap ($p < 0.05$): RI-MGN, H-MGN, and RG-MGN further increased correlation and reduced nRMSE to lower levels with markedly reduced variability. Differences among the three were relatively small, but overall, H-MGN maintained a high correlation while achieving lower nRMSE, reflecting a favorable trade-off between accuracy and consistency.

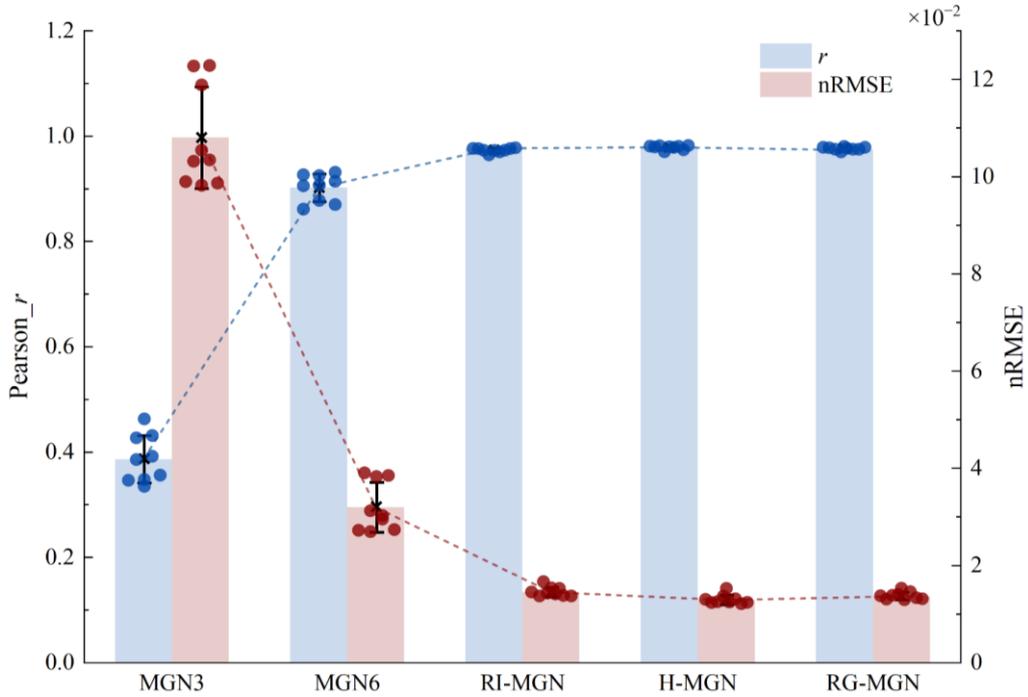

Figure 2. Consistency and normalized error in whole-field stress reconstruction across models. Solid circles represent individual subjects and are horizontally jittered around the corresponding bar center for clarity; the x-position has no statistical meaning. nRMSE and $r$ are shown in red and blue, respectively.

## 3.2 Reconstruction Accuracy for Peak and High-Quantile Stress

As shown in Figure 3, MGN3 and MGN6 exhibited larger error and dispersion in $RE_{max}$. Using only stacked local message passing to represent the amplitude of extreme stress-concentration regions was not stable. MGN6 improved compared with MGN3, but still could not reliably reproduce peak levels. In contrast, all three models with explicit region-interaction structures achieved significantly more accurate reconstruction of the high-stress tail ($p < 0.05$): $RE_{P95}$ was compressed to very low levels with markedly reduced variability. For $RE_{max}$, region-interaction models still showed distinguishable performance stratification: H-MGN achieved lower peak error and a more stable distribution, whereas RI-MGN and RG-MGN still displayed more pronounced peak deviations for some participants.

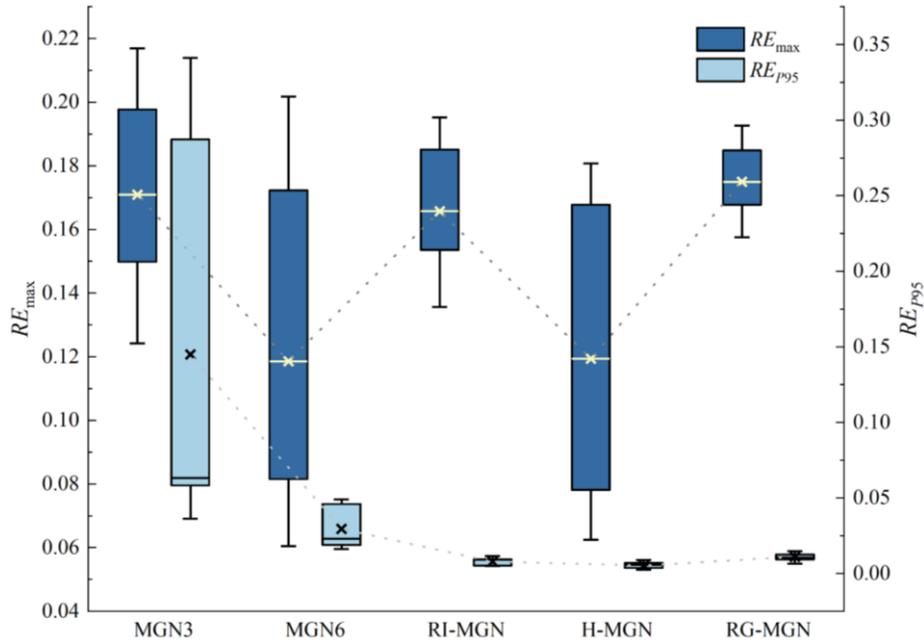

Figure 3. Comparison of errors in reconstructing peak and high-percentile stress across models. In each boxplot, the horizontal line indicates the median, the box indicates the interquartile range, the whiskers indicate the range, and × denotes the mean. Dotted lines connect the group means for visual guidance only.

## 3.3 High-Risk Region Consistency and Temporal Evolution of High-Quantile Error

As shown in Figure 4, models differed significantly in spatial consistency for reconstructing high-risk regions. MGN3 had the lowest Dice and IoU scores, with substantial inter-subject variability. MGN6 improved consistency, but still displayed substantial fluctuations. In contrast, all three explicit region-interaction models elevated Dice/IoU to high levels and showed significantly reduced dispersion ($p < 0.05$). Notably, subtle but stable differences were observed among region-interaction models: H-MGN and RI-MGN showed higher Dice/IoU with tighter distributions, indicating more stable hotspot-region reconstruction. Although RG-MGN also significantly outperformed both MGN-Stack baselines, it showed slight consistency drops in some participants.

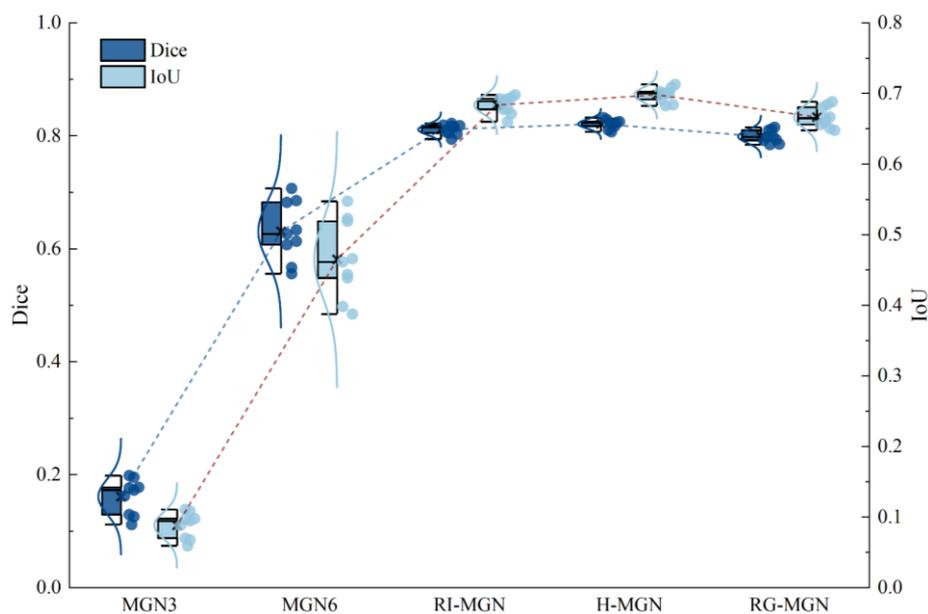

Figure 4. Comparison of spatial consistency for reconstructed high-risk regions across models. Solid dots represent individual subjects; the overlaid normal distribution curves summarize the across-subject distribution; × denotes the mean; the box indicates the interquartile range, and the line inside the box marks the median.

As shown in Figure 5, $RE_{P95}$ exhibited clear temporal heterogeneity during the stance phase. Overall, MGN3 maintained the highest $RE_{P95}$ and the widest fluctuation band. Although MGN6 significantly reduced error relative to MGN3, it remained at a noticeably higher level for most phases and showed more pronounced increases in error during the mid-to-late stance phase. In contrast, the three explicit region-interaction models compressed $RE_{P95}$ to near zero across all stance phases, with small fluctuations, capturing the temporal exposure trajectory of the high-stress tail. RI-MGN and H-MGN curves overlapped almost completely and remained the most stable throughout, indicating the greatest phase robustness. RG-MGN showed the same overall trend but exhibited a slight upturn in late stance.

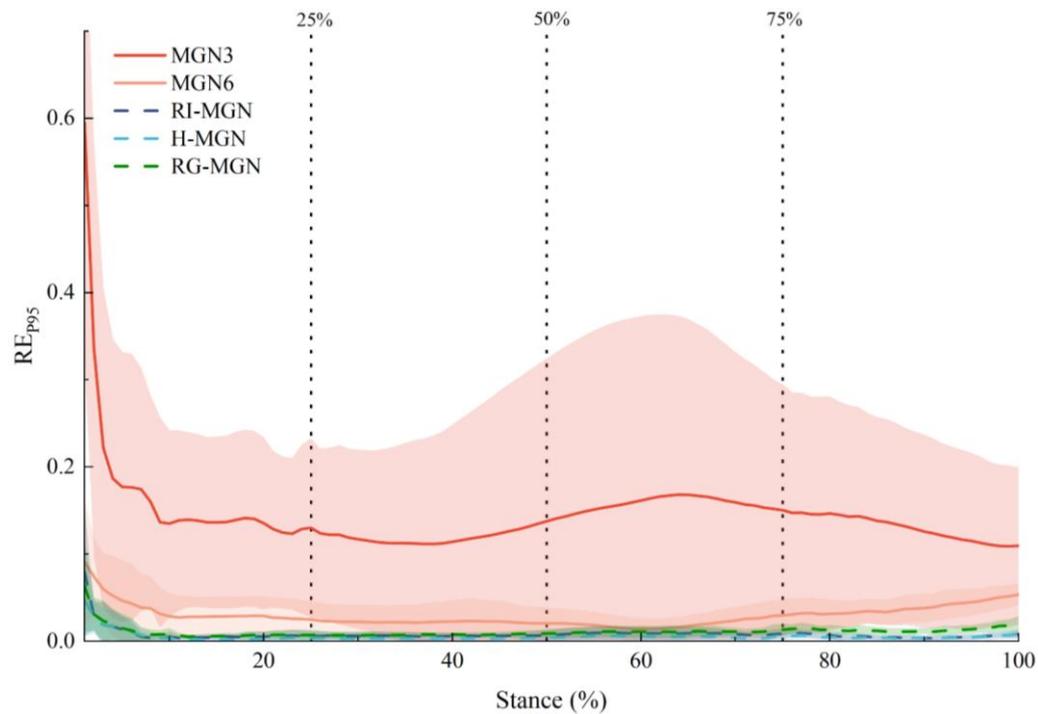

Figure 5. Temporal evolution of $RE_{P95}$ during stance. Lines show the mean across subjects, and the shaded bands indicate ± SD.

### 3.4 Hotspot Localization

Temporal trends in hotspot localization error are shown in Figure 6. MGN3 showed the largest deviation in hotspot localization throughout stance, with the widest fluctuations. Although MGN6 reduced localization error relative to MGN3, it remained relatively high in early and mid-stance, and its error was more sensitive to phase changes. In contrast, RI-MGN, H-MGN, and RG-MGN maintained very low and stable hotspot distances across all phases, with markedly reduced variability. Differences among these three were small: H-MGN and RI-MGN had smoother curves and narrower error bands overall, reflecting more consistent hotspot localization, while RG-MGN still significantly outperformed MGN-Stack baselines but showed slight increases at a few phases.

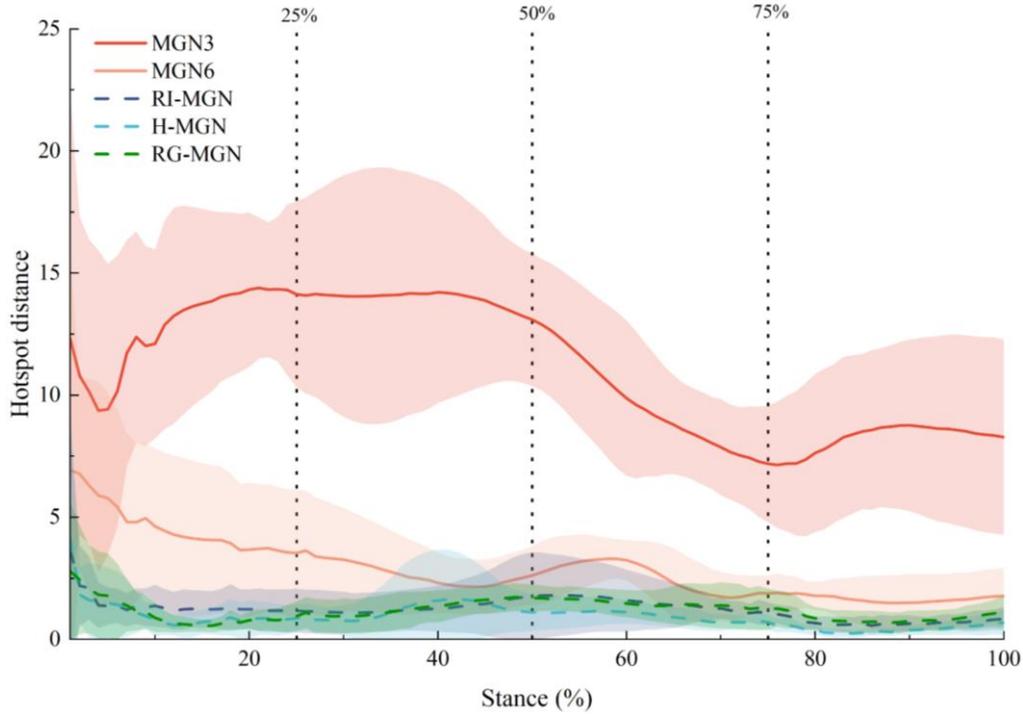

Figure 6. Temporal evolution of hotspot localization error during stance (mean ± SD).

## 4 Discussion

This study systematically compared pure stacked local propagation with three explicit region-interaction architectures for reconstructing knee contact stress fields across subjects. We found that stacking local message passing can improve whole-field reconstruction accuracy, but the gains are clearly limited on hotspot-relevant metrics. In contrast, explicitly introducing region-level interactions produced more pronounced and more stable improvements in high-risk-region consistency and in controlling high-quantile errors. These findings suggest that long-range dependencies in multi-tissue coupled knee contact are not primarily achieved through deeper or more repeated local diffusion; rather, they are more likely determined by coupling among a small number of load-bearing/contact-relevant regions.

Existing MGN/MP-GNN surrogate models often treat joint contact mechanics as a local message-passing regression problem on a fixed mesh topology, expanding the receptive field by increasing network depth or the number of propagation steps [6]. However, in knee multi-tissue coupled contact, key responses often concentrate around load-bearing regions and contact pathways [18]. Far-field effects are not simple uniform diffusion; they behave more like directed transfer among a few key regions. Therefore, while stacking local propagation can partially mitigate insufficient long-range modeling, multi-hop diffusion along the mesh topology can lead to over-smoothing and noise accumulation [19], making it difficult for distant influences to reach hotspot regions with a high signal-to-noise ratio. This helps explain why stacked local propagation lags behind region-interaction architectures. The findings also highlight the benefit of elevating the structural representation unit from elements/nodes to regions/hierarchies in joint-contact-mechanics surrogates.

Mechanistically, the advantages of RI-MGN and H-MGN can be understood as shifting long-range dependency modeling from full-mesh diffusion to interactions among a small number of regional entities: RI-MGN aggregates cross-region information through a low-rank token channel; H-MGN performs multi-scale pool–interact–unpool, integrating long-range dependencies at a coarse scale before feeding them back to the fine scale, making it easier to form stable cross-region coupling representations. RG-MGN uses gating to directly apply region information to the message-passing gain. It significantly outperforms the pure stacking baseline while keeping a compact structure. Still, it shows mild fluctuations in some phases, suggesting that gating-based inductive bias may be somewhat less adaptive to fine-grained hotspot migration. Overall, the common advantage of the three region-interaction models is that they achieve greater fidelity to hotspot-related features with fewer local propagation steps, reflecting a better accuracy–efficiency balance.

From a computational and engineering perspective, the comparison also yields a practical implication: under the same GPU memory budget and similar training configurations, region-interaction models can achieve better hotspot-relevant performance with fewer local propagation steps, demonstrating a superior accuracy–efficiency tradeoff. For high-resolution mesh-based biomechanical field prediction, simply increasing depth/width often leads to nonlinear growth in memory and computational costs [20]. In contrast, region interactions move long-range modeling into lower-dimensional channels, offering a feasible route to improving hotspot fidelity under resource constraints. This is especially important for future training and inference on larger samples and more complex movement libraries. As data and mesh scales continue to grow, structural improvements are often more sustainable than "throwing more compute" at the problem.

Several limitations should be noted. First, the sample size and population diversity are limited, and the task scenarios focus on a specific movement pattern, which may limit the extrapolation of the conclusions to broader populations and movement libraries. Second, the finite element simulations were quasi-static and did not explicitly include time-dependent factors such as loading history, so the current findings primarily concern representation of instantaneous spatial long-range dependencies; true temporal memory effects require further validation. Finally, key hyperparameters in region-interaction models remain adjustable; future work should systematically analyze their effects on stability and interpretability under fair computational budgets.

## 5 Conclusion

In knee contact stress-field prediction, compared with simply stacking local message passing, explicitly introducing region-level interaction mechanisms more stably improves fidelity in the high-stress tail and hotspot localization, and markedly enhances reconstruction consistency and phase robustness for high-risk regions. Overall, region interactions provide a more efficient modeling path that is more aligned with the structure of mechanical stress propagation, enabling more reliable recovery of risk-relevant stress phenotypes under limited computational budget, and providing structural support for developing interpretable stress phenotypes and mechanical surrogate models for injury-risk assessment.

## Funding Statement

This study was supported by the National Key R&D Program of China (grant no. 2023YFC2411201); Beijing Natural Science Foundation (grant no. L259081); NSFC General Program (grant no. 31870942); Peking University Clinical Medicine Plus X – Young Scholars Project (grant nos. PKU2020LCXQ017 and PKU2021LCXQ028); and PKU-Baidu Fund (grant no. 2020BD039).

## Declaration of Interest Statement

The authors have no conflicts of interest to declare.

Networks. Advances in Neural Information Processing Systems (NeurIPS 2021), 2021.